# Resilience-Motivated Distribution System Restoration Considering Electricity-Water-Gas Interdependency

Jiaxu Li, *Graduate Student Member, IEEE,* Yin Xu, *Senior Member, IEEE,* Ying Wang, *Member, IEEE,* Meng Li, *Member, IEEE,* Jinghan He, *Fellow, IEEE,* Chen-Ching Liu, *Life Fellow, IEEE,* and Kevin P. Schneider, *Fellow, IEEE*

*Abstract*—A major outage in the electricity distribution system may affect the operation of water and natural gas supply systems, leading to an interruption of multiple services to critical customers. Therefore, enhancing resilience of critical infrastructures requires joint efforts of multiple sectors. In this paper, a distribution system service restoration method considering the electricity-water-gas interdependency is proposed. The objective is to provide electricity, water, and natural gas supplies to critical customers in the desired ratio according to their needs after an extreme event. The operational constraints of electricity, water, and natural gas networks are considered. The characteristics of electricity-driven coupling components, including water pumps and gas compressors, are also modeled. Relaxation techniques are applied to nonconvex constraints posed by physical laws of those networks. Consequently, the restoration problem is formulated as a mixed-integer second-order cone program, which can readily be solved by the off-the-shelf solvers. The proposed method is validated by numerical simulations on electricity-water-gas integrated systems, developed based on benchmark models of the subsystems. The results indicate that considering the interdependency refines the allocation of limited generation resources and demonstrate the exactness of the proposed convex relaxation.

## I. Nomenclature

### A. Notation Associated with Electricity Distribution System

*Sets and Parameters*

| | |
|---|---|
| $\mathcal{N}^{E}$ | Set of nodes in electricity distribution system, $\mathcal{N}^{E} = \mathcal{N}^{L} \cup \mathcal{N}^{pump} \cup \mathcal{N}^{comp} \cup \mathcal{N}^{DG}$. $|\mathcal{N}^{E}|$ is the number of nodes |
| $\mathcal{N}^{L}, \mathcal{N}^{pump},$ $\mathcal{N}^{comp}, \mathcal{N}^{DG}$ | Set of nodes with customers, water pumps, gas compressors, and DGs, respectively |
| $\mathcal{E}^{E}$ | Set of branches in electricity distribution system |
| $\mathcal{R}$ | Set of the root nodes providing fictitious flow. In this paper, only one root node is defined to |

avoid multiple islands after restoration, i.e., $|\mathcal{R}| = 1$

| | |
|---|---|
| $a_i, b_i, c_i$ | Weighting factors of electricity, water, and natural gas demand of customer at node $i$. |
| $D_i$ | Fictitious demand at node $i$, which can be set as 1 for each non-root node |
| $\overline{l_{ij}}$ | Upper limit of current magnitude squared on branch $i \rightarrow j$ |
| $s_i^{rate}$ | Electricity demand of customer at node $i$ |
| $\overline{S_i^{DG}}$ | Upper limits of DG apparent power generation at node $i$ |
| $\underline{v_i}, \overline{v_i}$ | Lower and upper limits of voltage magnitude squared at node $i$ |
| $\omega_i$ | Priority weight associated with customer at node $i$. |
| $\lambda$ | Weighting coefficients of penalty term in objective |
| $\varphi_i^{pump}, \varphi_i^{comp}$ | Power factor of the water pump or gas compressor at node $i$, respectively |

*Variables*

| | |
|---|---|
| $a_{ij}$ | Restoration status of branch $i \rightarrow j$, a binary variable. If branch $i \rightarrow j$ is connected, $a_{ij} = 1$; otherwise, $a_{ij} = 0$ |
| $F_{ij}$ | Fictitious flow on branch $i \rightarrow j$ |
| $l_{ij}$ | Current magnitude squared on branch $i \rightarrow j$ |
| $P_i^{pump}$ | Real power consumption of pump at node $i$ |
| $P_i^{comp}$ | Real power consumption of compressor at node $i$ |
| $r_i$ | Electricity demand restoration status of customer at node $i$, a binary variable. If customer at node $i$ is restored, $r_i = 1$; otherwise, $r_i = 0$ |
| $s_i$ | Complex power injection at node $i$ |
| $s_i^{DG}$ | Complex power output of DG at node $i$ |
| $S_{ij}$ | Sending-end complex power flow on branch $i \rightarrow j$ |
| $v_i$ | Voltage magnitude squared at node $i$ |

This work was supported by the Fundamental Research Funds for the Central Universities under Grant No. 2018JBZ004. (Corresponding Author: Meng Li)

J. Li, Y. Xu, Y. Wang, M. Li, and J. He are with the School of Electrical Engineering, Beijing Jiaotong University, Beijing 100044, China (e-mails: lijiaxu@bjtu.edu.cn, xuyin@bjtu.edu.cn, yingwang1992@bjtu.edu.cn, mengl@bjtu.edu.cn, jhhe@bjtu.edu.cn).

C.-C. Liu is with The Bradley Department of Electrical and Computer Engineering, Virginia Polytechnic Institute and State University (VT), Blacksburg, VA 24061, USA (e-mail: ccliu@vt.edu).

K. P. Schneider is with the Pacific Northwest National Laboratory (PNNL) located at the Battelle Seattle Research Center in Seattle, Washington. (e-mail: kevin.schneider@pnnl.gov).

$\chi_i$     ON/OFF status of the water pump or gas compressor at node $i$, a binary variable. If it is ON, $\chi_i = 1$; otherwise, $\chi_i = 0$

*B. Notation Associated with Water Supply System*

*Sets and Parameters*

| | |
|---|---|
| $\mathcal{N}^{\mathrm{W}}$ | Set of nodes in water supply system |
| $\mathcal{E}^{\mathrm{W}}$ | Set of branches in water supply system, and $\mathcal{E}^{\mathrm{W}} = \mathcal{E}_{\mathrm{W}}^{\mathrm{pipe}} \cup \mathcal{E}_{\mathrm{W}}^{\mathrm{pump}}$ |
| $\mathcal{E}_{\mathrm{W}}^{\mathrm{pipe}}, \mathcal{E}_{\mathrm{W}}^{\mathrm{pump}}$ | Set of pipes and pumps in water supply system, respectively |
| $g^{\mathrm{W}}$ | Standard gravity. $g^{\mathrm{W}} = 9.8 \mathrm{m/s^2}$. |
| $\underline{h_i}$ | Lower limit of water head at node $i$ |
| $\overline{w_i^{\mathrm{rate}}}$ | Water demand of customer at node $i$ |
| $\overline{W_{ij}}$ | Upper limit of water flow on branch $i \to j$ |
| $\alpha, \beta$ | Water pump parameters |
| $\eta^{\mathrm{pump}}$ | Water pump efficiency |
| $\rho^{\mathrm{W}}$ | Water density. $\rho^{\mathrm{W}} = 10^3 \mathrm{kg/m^3}$. |

*Variables:*

| | |
|---|---|
| $w_i$ | Water demand restoration status of customer at node $i$ |
| $w_i^{\mathrm{WR}}$ | Water output from water reservoir at node $i$ |
| $W_{ij}$ | Water flow on branch $i \to j$ |
| $h_i$ | Water head at node $i$ |
| $\Delta h_{ij}$ | Water head loss on branch $i \to j$ |

*C. Notation Associated with Natural Gas Supply System*

Sets and Parameters

| | |
|---|---|
| $\mathcal{N}^{\mathrm{G}}$ | Set of nodes in natural gas supply system |
| $\mathcal{E}^{\mathrm{G}}$ | Set of branches in natural gas supply system, and $\mathcal{E}^{\mathrm{G}} = \mathcal{E}_{\mathrm{G}}^{\mathrm{pipe}} \cup \mathcal{E}_{\mathrm{G}}^{\mathrm{comp}}$ |
| $\mathcal{E}_{\mathrm{G}}^{\mathrm{pipe}}, \mathcal{E}_{\mathrm{G}}^{\mathrm{comp}}$ | Set of pipes and compressors in natural gas supply system, respectively |
| $g_i^{\mathrm{rate}}$ | Natural gas demand of customer at node $i$ |
| $\gamma$ | Compression ratio of compressor |
| $\sigma$ | Conversion rate of compressor |
| $\eta^{\mathrm{comp}}$ | Compressor efficiency |

*Variables:*

| | |
|---|---|
| $g_i$ | Natural gas demand restoration status of customer at node $i$ |
| $g_i^{\mathrm{GS}}$ | Natural gas output from gas station at node $i$ |
| $G_{ij}$ | Gas flow on branch $i \to j$ |
| $\psi_i$ | Gas pressure magnitude squared at node $i$ |
| $\Delta \psi_{ij}$ | Gas pressure magnitude squared drop on branch $i \to j$ |

## II. INTRODUCTION

DUE to the increasing number of major power outages caused by natural disasters, cyber-attacks, and accidents, power system resilience has received much attention by the power industry and researchers [1], [2]. After a major outage, using locally available power sources to restore critical loads can improve resilience and reduce economic losses [3]. Local sources used for service restoration include distributed generators (DGs) [4], energy storage (ES) [5], intermittent energy resources [6], electrical vehicles (EVs) [7], and mobile emergency resources

(MERs) [8], etc. Microgrids and networked microgrids can also be used as a resiliency resource to restore critical loads on utility feeders [9], [10]. During restoration, the distribution network can be sectionalized into several small electrical islands [4]-[7],[9] or a large island can be formed to share limited generation resources in a wider range [3], [8], [10].

A major outage in the electricity distribution system may also interrupt the water and natural gas supplies to customers. In a water supply system, water treatments, which consume electricity, produce potable water for urban customers [11]. Electricity-driven water pumps are employed to boost the water head, compensating water head losses along a water pipe [12]. In a natural gas supply system, gas compressors run by electric motors are used to maintain an acceptable gas pressure at nodes [13]. The electricity demand of water treatments, water pumps, and gas compressors depends on the service area [14], [15]. These facilities would not work properly when the electricity supply is interrupted, which further affects water and natural gas supplies to customers.

Aware of such threats, researchers have paid growing attention to the electricity-water-gas interdependency in deciding the restoration strategy. Decision-making methods considering the interdependency fall into three categories. The first approach is based on graph theory. The electricity, water, and natural gas systems are represented as networks (or graphs) while the operational characteristics are not considered. The interdependency is described by the connection among multiple networks [16], [17]. The second approach regards water pumps and gas compressors as critical loads [18], while the operational constraints of the water and natural gas supply systems are not considered. Ref. [19] proposes a service restoration method considering the operational capacity of a hospital and a water pump. The water delivered to the hospital is described by the power available to the water pump. However, it cannot be applied to the system with multiple critical loads and a complex water supply system. The third approach considers the operational characteristics of electricity, water, and natural gas supply systems. These studies are mainly conducted on the electricity-gas integrated system. Ref. [20] focuses on the repair crew dispatch problem. The status of damaged component, including distribution lines, DGs, gas pipelines, and gas compressors, are correlated with the sequential repair crew dispatch. Ref. [21] focuses on the conversion between electricity and natural gas. The two systems can support each other in the restoration process through gas-fired DGs and power-to-gas devices.

To the best of our knowledge, few studies have focused on the electricity, water, and natural gas demand of customers when determining service restoration strategies. The customers need a proper amount of electricity, water, and natural gas to perform critical tasks [22], [23]. In addition, delivering water and natural gas to customers relies on electricity. However, the existing methods mentioned above cannot ensure optimal allocation of limited generation power capacity to satisfy demand of end-users. For the first approach, the obtained restoration strategy may violate the operational constraints of the electricity, water, or natural gas supply systems. The second approach usually allocates generation power by assigning priority to critical loads. Since it is widely used in service restoration problems, its limitation will be discussed and illustrated with an example in the next section. The third approach is still relatively preliminary and needs further research. The method in this paper is developed based on the third approach.

In this paper, a service restoration method for electricity distribution systems considering the electricity-water-gas interdependency is proposed. The electricity, water, and natural gas demand of customers can be restored coordinately based on their needs. The operational characteristic of electricity, water, and natural gas networks are considered. The major contributions include:

1) An optimization model is proposed for the service restoration problem, incorporating the operational constraints of the electricity, water, and natural gas supply networks and the coupling among them. It is formulated as a mixed-integer nonlinear program (MINLP).

2) A convexification method is proposed to reformulate the MINLP as a mixed-integer second-order cone program (MISOCP) using relaxation techniques. The operational constraints of the electricity, water, and natural gas networks, as well as the electricity consumption constraints of water pumps and gas compressors, are relaxed as second-order cone constraints. The relaxation is proved to be exact under certain conditions. The resulted MISOCP can be solved efficiently to meet the online decision-making requirements.

3) The benefits of considering interdependency in the service restoration model are analyzed and demonstrated in case studies.

Noted that two test systems integrating electricity, water, and natural gas networks are developed for the first time based on the standard test systems provided in multiple disciplines, which can be used by peers who are interested in the research related to the interdependency of the three networks. This is also a contribution of this paper.

The remainder of this paper is organized as follows. Section III describes the service restoration problem considering interdependency and its challenges. The service restoration model is proposed in Section IV. Section V presents the convex relaxation methods converting the MINLP to a MISOCP and illustrates the exactness conditions. Case studies are presented in Section VI. Section VII concludes the paper.

## III. PROBLEM DESCRIPTION

The operation of electricity, water, and natural gas supply systems relies on each other in multiple scales through various components. This paper focuses on the electricity needs for moving water and natural gas from reservoirs to customers in the distribution scale. The structure of the electricity-water-gas integrated system considered in this paper is shown in Fig. 1. The system consists of an electricity distribution system with distributed energy resources, a water supply system with pumps, and a natural gas supply system with compressors. It is assumed that the topologies of water and natural gas systems are radial.

The water pumps overcome elevation differences between nodes and water head losses in pipes to move water. Gas compressors are placed on selected pipes to increase the pressure at their output node. Both water pumps and gas compressors require electricity to operate. Therefore, the operation of water and natural gas supply systems relies on the electric network. It is also assumed that customers in this system model demand electricity, water, and natural gas.

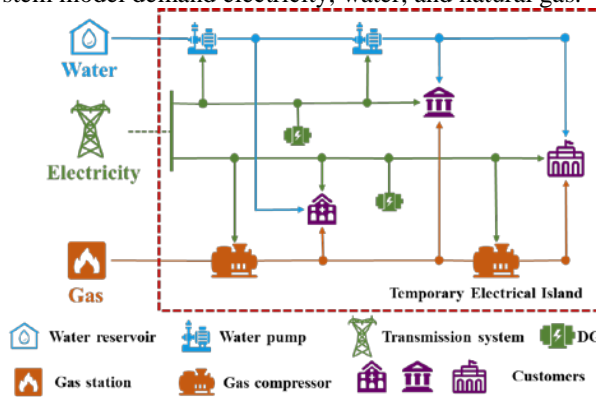

Fig. 1. Structure of the electricity-water-gas integrated system.

The major outage scenario is considered. The purpose of this paper is to develop a service restoration method for electricity distribution systems aiming at providing electricity, water, and natural gas to critical customers. The following assumptions are made:

1) The utility power is not available. It may take several hours or even days to repair damaged components and restore the transmission system.

2) Due to power outage, water pumps and gas compressors are shut off. Although there may be backup power, it can usually operate for just a few hours due to limited fuel [14]. After that, the water and natural gas supplies to customers are also interrupted.

3) Local generation sources are coordinated to form a temporary electrical island in the distribution level to supply electricity to the loads, including end-users, electricity-driven water pumps, and gas compressors run by electric motors.

4) This paper is focused on deciding the post-restoration state, including the post-restoration topology of the temporary electrical island, real and reactive power outputs of DGs, satisfied demand of end-users, and the electric power supplied to water pumps and gas compressors. The major task is properly allocating limited generation resources to satisfy the electricity, water, and natural gas demand of end-users. The sequential restoration actions are not within the scope.

In service restoration, limited generation resources are allocated to critical loads usually by assigning priority to loads, including end-users and electricity-driven facilities in water and natural gas supply systems, regardless of the operational characteristics of the two systems. However, a fixed priority setting cannot fit varying operating conditions. In some cases, even for a given operating condition, no suitable priority setting can guarantee a proper allocation of generation resources that satisfies electricity, water, and natural gas demand of end-users. For example, assume that two pumps with the same electric capacity, 400kW, are located in a water distribution path in series, as shown in Fig. 2. The water demand of two customers is 250 m³/h and 280 m³/h, respectively. The total available electric power is 400kW. By assigning priority to the two pumps, the objective of service restoration can be:

1) Maximizing the weighted electric power supply to water pumps [5], formulated by (1).
$$\max \omega_1 P_1 + \omega_2 P_2 \qquad (1)$$
where $\omega_1$ and $\omega_2$ are priority weights of the two water pumps. $P_1$ and $P_2$ are the electric power supply to water pumps.

2) Maximizing the weighted number of restored water pumps [3], formulated by (2).
$$\max \omega_1 x_1 + \omega_2 x_2 \qquad (2)$$
where $x_1$ and $x_2$ are binary variables representing the restoration status of the two water pumps. If pump $i$ is restored, $x_i = 1$; otherwise, $x_i = 0$, where $i = 1,2$.

3) Maximizing the weighted percentage of load (electric power) restored at each water pump [24], formulated by (3).
$$\max \omega_1 z_1 + \omega_2 z_2 \qquad (3)$$
where $z_1$ and $z_2$ denote the percentage of load restored at each water pump, i.e., $z_i = P_i / P_i^{\max}$ ($i = 1,2$), and $P_i^{\max}$ means the electric capacity of pump $i$.

The possible post-restoration states of the water supply system are illustrated as below:

1) If $\omega_1 > \omega_2$, the upstream water pump is restored with full capacity due to the higher priority while the other receives no power. Thus, only the water demand of customer 1 can be satisfied, as shown in Fig. 2 (a).

2) If $\omega_1 < \omega_2$, the downstream water pump is restored with full capacity due to the higher priority while the upstream one receives no power. As a result, water cannot reach any customer, as shown in Fig. 2 (b).

3) If $\omega_1 = \omega_2$, both pumps can be partially restored with $P_1 + P_2 = 400$ kW, as shown in Fig. 2 (c). In this case, one cannot determine an optimal restoration strategy only based on the priority setting, because the objective value does not change with the power allocation strategy. In this case, more factors such as the topology and operational characteristics of the water supply system need to be considered.

In summary, no matter how the priority is assigned, optimal allocation of generation power is not guaranteed. Therefore, it is necessary to consider the water demand of customers and the operational characteristics of the water supply system to obtain an efficient restoration strategy, as shown in Fig. 2 (d), especially for more complicated cases with a relatively large water supply network and multiple water pumps.

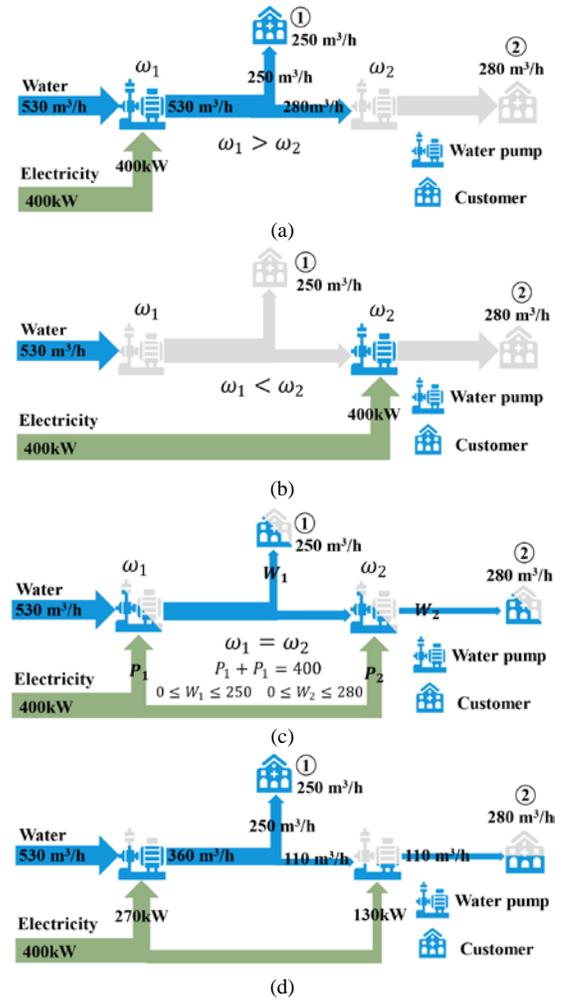

Fig. 2. Post-restoration states of a water supply system. (a)-(c) are determined by assigning priority to water pumps with (a) $\omega_1 > \omega_2$, (b) $\omega_1 < \omega_2$, and (c) $\omega_1 = \omega_2$. (d) is determined according to the customers' demand and operational characteristics of the water supply system.

The major challenge of allocating limited generation resources lies in the conflict between the heavy computational burden and the online decision-making requirement. The operational constraints of the electricity, water, and natural gas networks must be satisfied. These constraints are described by a set of nonlinear and nonconvex equations, including the electric power flow equations [25], water head loss equations [12], and gas pressure drop equations [13]. In addition, the electricity consumption constraints of water pumps and gas compressors contain logic propositions. The resulted optimization model is a nonlinear and nonconvex model with many integer variables, making it hard to solve. However, the service restoration strategy should be determined online after the prior-fault operating condition and fault locations are known. Multiple faults may occur after major outages and the prior-fault operating condition varies in a large range. Therefore, it is not practical to identify critical scenarios and decide candidate restoration plans offline. In this case, a computationally efficient and reliable decision-making method is required.

## IV. PROBLEM FORMULATION

In this section, the proposed optimization model for the service restoration problem is presented.

### A. Decision Variables

The decision variables contain restoration status of, customers and coupling components, as well as DG outputs. For customers, the status of electricity demand $r_i$ is discrete, while the water $w_i$ and natural gas $g_i$ are continuous. For coupling components, the restoration status $\chi_i$ represents ON/OFF. When they are switched ON, the electricity consumptions of water pumps $P_i^{\text{pump}}$ and gas compressors $P_i^{\text{comp}}$ are variable.

### B. Objective Function

The purpose of service restoration is to provide electricity, water, and natural gas supplies to critical customers after a major outage. A weighting factor $\omega_i > 0$ is assigned to each customer based on their priority. A larger value of $\omega_i$ indicates higher priority. At the meantime, for a restored customer, certain amount of electricity, water, and natural gas should be provided in a ratio desired by the customer. The ratio reflects how heavy the customer's critical function depends on each type of resources, relatively. Parameters $a_i, b_i, c_i \geq 0$ (with $a_i + b_i + c_i = 1$) are used to represent the ratio. A relatively larger value of $a_i$ ($b_i$ or $c_i$) indicates this customer relies more on electricity (water or gas) rather than the other two resources.

Based on the above analysis, the optimization objective can be designed as follows:

$$\max f = \sum_{i=1}^{N^{\text{L}}} \omega_i \min \left\{ \frac{r_i}{a_i}, \frac{w_i/w_i^{\max}}{b_i}, \frac{g_i/g_i^{\max}}{c_i} \right\}, i \in \mathcal{N}^{\text{L}} \quad (4)$$

where $r_i$, $w_i/w_i^{\max}$, and $g_i/g_i^{\max}$ denote the normalized amount of electricity, water, and natural gas demand restored, respectively. It is assumed that electricity demand can only be fully restored or not restored, because the electricity load is controlled by a switch (either ON or OFF). In contrast, the water/gas demand can be partially restored by controlling the water/gas pressure continuously. Therefore, $r_i \in \{0,1\}$, while $w_i/w_i^{\max}$ and $g_i/g_i^{\max} \in [0,1]$.

To see that the objective given in (4) serves the purpose of allocating resources in the ratio desired by the customer, consider a customer $i$ to be restored. Assume that the ratio of its desired amount of electricity, water, and gas (normalized) is $a_i : b_i : c_i$. Since only the minimum of $\frac{r_i}{a_i}, \frac{w_i/w_i^{\max}}{b_i}$, and $\frac{g_i/g_i^{\max}}{c_i}$ contributes the objective value, the optimal solution tends to have the ratio $r_i : (w_i/w_i^{\max}) : (g_i/g_i^{\max})$ as close to $a_i : b_i : c_i$ as possible. If not, it means some resources are allocated to customer $i$ but do not contribute to the objective value. These portions of resources can be reallocated to other customers to further increase the objective value, which contradicts the assumption that it is the optimal solution.

### C. Constraints

#### 1) Operational Constraints on Electricity Distribution System

The operational constraints on electricity distribution system can be expressed as follows [26].

$$\sum_{k:k \to i} (S_{ki} - z_{ki} l_{ki}) + s_i = \sum_{j:i \to j} S_{ij}, i \in \mathcal{N}^{\text{E}} \quad (5)$$

$$s_i = s_i^{\text{DG}} - r_i s_i^{\text{rate}}, i \in \mathcal{N}^{\text{DG}} \quad (6)$$

$$s_i = -r_i s_i^{\text{rate}}, i \in \mathcal{N}^{\text{L}} \quad (7)$$

$$v_i - v_j \leq M(1 - a_{ij}) + 2\text{Re}(z_{ij}^{\text{H}} S_{ij}) - |z_{ij}|^2, \\ i \to j \in \mathcal{N}^{\text{E}} \quad (8)$$

$$v_i - v_j \geq -M(1 - a_{ij}) + 2\text{Re}(z_{ij}^{\text{H}} S_{ij}) - |z_{ij}|^2, \\ i \to j \in \mathcal{N}^{\text{E}} \quad (9)$$

$$|S_{ij}|^2 = v_i l_{ij}, i \to j \in \mathcal{E}^{\text{E}} \quad (10)$$

$$0 \leq |s_i^{\text{DG}}| \leq \overline{S_i^{\text{DG}}}, i \in \mathcal{N}^{\text{DG}} \quad (11)$$

$$\underline{v_i} \leq v_i \leq \overline{v_i}, i \in \mathcal{N}^{\text{E}} \quad (12)$$

$$l_{ij} \leq a_{ij} \overline{l_{ij}}, i \to j \in \mathcal{E}^{\text{E}} \quad (13)$$

$$\sum_{i \to j \in \mathcal{E}^{\text{E}}} a_{ij} = |\mathcal{N}^{\text{E}}| - 1, i \to j \in \mathcal{E}^{\text{E}} \quad (14)$$

$$\sum_{k:k \to i} F_{ki} = \sum_{j:i \to j} F_{ij} + D_i, i \in \mathcal{N}^{\text{E}} \backslash \mathcal{R} \quad (15)$$

$$|F_{ij}| \leq a_{ij} M, i \to j \in \mathcal{E}^{\text{E}} \quad (16)$$

Constraints (5)-(10) represent the power flow constraint [27]. This paper considers the three-phase balanced power flow, because water pumps and gas compressors driven by three-phase power are widely used [28]. Constraint (11) indicates that the DG outputs cannot exceed their capacities. Constraints (12) and (13) ensure the node voltage and branch current are within the limits. Constraints (14)-(16) are radial topology constraints based on single-commodity flow method [29]. It is assumed that fictitious flow exists in the topology and one root node serves as the fictitious source while other nodes have fictitious demand. Constraint (14) means the number of connected lines equals the number of nodes minus 1. Constraint (15) requires that fictitious flow balance is satisfied for each non-root node. Constraint (16) forces $F_{ij}$ to zero when the line is disconnected. Constraints (15)-(16) ensure that all other nodes are connected to the root node in order to satisfy the fictitious load demand.

#### 2) Operational Constraints on Water Supply System

The operational constraints on water supply system are described in detail as follows [30], [31].

$$\sum_{k:k \to i} W_{ki} + (w_i^{\text{WR}} - w_i) = \sum_{j:i \to j} W_{ij}, i \in \mathcal{N}^{\text{W}} \quad (17)$$

$$0 \leq w_i \leq w_i^{\text{rate}}, i \in \mathcal{N}^{\text{W}} \quad (18)$$

$$\Delta h_{ij} = h_i - h_j, i \to j \in \mathcal{E}^{\text{W}} \quad (19)$$

$$\Delta h_{ij} = F_{ij} W_{ij}^2, i \to j \in \mathcal{E}_{\text{W}}^{\text{pipe}} \quad (20)$$

$$\begin{cases} \Delta h_{ij} = -(\alpha W_{ij} + \beta) & \text{if } \chi_m = 1 \\ W_{ij} = 0 & \text{if } \chi_m = 0 \end{cases} \\ i \to j \in \mathcal{E}_{\text{W}}^{\text{pump}}, m \in \mathcal{N}^{\text{pump}} \quad (21)$$

$$0 \leq W_{ij} \leq \overline{W_{ij}}, i \to j \in \mathcal{E}^{\text{W}} \quad (22)$$

$$\underline{h_i} \leq h_i, i \in \mathcal{N}^{\text{W}} \quad (23)$$

Constraint (17) represents the water flow balance at each node. The amount of restored water demand is limited by (18). Constraint (19) demonstrates the water head variation between adjacent nodes. Constraint (20) is Darcy–Weisbach equation that represents the water head losses in pipes, where $F_{ij}$ is a constant [30]. Constraint (21) represents the hydraulic characteristics of pumps. If the pump is restored, the relationship between water head gain and water flow through the pump is described as a linear expression [30]. Constraints (22) and (23) indicate that the water flow on branch and water head at each node are bounded.

*3) Operational Constraints on Natural Gas Supply System*

The operational constraints on natural gas supply system are described as [32]:

$$\sum_{k:k\to i} G_{ki} + (g_i^{GS} - g_i) = \sum_{j:i\to j} G_{ij}, i \in \mathcal{N}^G \quad (24)$$

$$0 \le g_i \le g_i^{rate}, i \in \mathcal{N}^G \quad (25)$$

$$\psi_i - \psi_j = \Delta\psi_{ij}, i \to j \in \mathcal{E}_W^{pipe} \quad (26)$$

$$C_{ij}\Delta\psi_{ij} = G_{ij}^2, i \to j \in \mathcal{E}_G^{pipe} \quad (27)$$

$$\begin{cases} \psi_i \le \psi_j \le \gamma\psi_i & if \ \chi_m = 1 \\ G_{ij} = 0 & if \ \chi_m = 0 \end{cases},$$
$$i \to j \in \mathcal{E}_G^{comp}, m \in \mathcal{N}^{comp} \quad (28)$$

$$0 \le G_{ij} \le \overline{G_{ij}}, i \to j \in \mathcal{E}^G \quad (29)$$

$$\underline{\psi_i} \le \psi_i \le \overline{\psi_i}, i \in \mathcal{N}^G \quad (30)$$

Constraint (24) represents the gas flow balance at each node. Constraint (25) means the restored natural gas demand should not exceed its rated capacity. Constraints (26) and (27) demonstrate the gas pressure drop in pipes based on Weymouth equation, where $C_{ij}$ is a constant. Constraint (28) is the operational constraint of gas compressors. If the compressor is restored, the node pressure at its output can be increased by a factor $\gamma$ [20]. Constraints (29) and (30) ensure that gas flow on each branch and gas pressure at each node stay within a preset range.

*4) Electricity Consumption for Water Pumps and Gas Compressors*

The electricity consumption for water pumps and gas compressors can be calculated as follows.

$$P_m^{pump} = \frac{\rho^W g^W}{\eta^{pump}}(\alpha W_{ij}^2 + \beta W_{ij}),$$
$$m \in \mathcal{N}^{pump}, i \to j \in \mathcal{E}_W^{pump} \quad (31)$$

$$P_m^{comp} = \sigma G_{ij}, m \in \mathcal{N}^{comp}, i \to j \in \mathcal{E}_G^{comp} \quad (32)$$

Constraint (31) links the electricity distribution system with the water supply system. The electricity consumption of water pumps is a quadratic function of the water flow [30]. Constraint (32) connects the electricity distribution system and natural gas supply system. The electricity consumption of a gas compressor is proportional to the gas flow [20].

The electricity-driven water pumps and gas compressors run by electric motors are considered as loads in the electric network. The power factor depends on the rotor speed. In this paper, a fixed value of power factor, i.e., 0.85 (lagging), is used for all water pumps and gas compressors. They are modeled as injected complex power at respective nodes, i.e.,

$$s_i = -\chi_i(P_i^{pump} + j\varphi_i^{pump}P_i^{pump}), i \in \mathcal{N}^{pump} \quad (33)$$

$$s_i = -\chi_i(P_i^{comp} + j\varphi_i^{comp}P_i^{comp}), i \in \mathcal{N}^{comp} \quad (34)$$

*D. Optimization Model*

To sum up, an MINLP model for the service restoration is obtained, which is referred to as **SRS-MINLP.**

**SRS-MINLP**

$$\max \quad (4)$$

over: $a_{ij}, F_{ij}, r_i, w_i, g_i, P_i^{pump}, P_i^{comp}, s_i^{DG}, S_{ij}, s_i,$
$\quad v_i, l_{ij}, \chi_i, w_i^G, W_{ij}, h_i, \Delta h_{ij}, g_i^G, G_{ij}, \psi_i, \Delta\psi_{ij}$

s.t. (5)-(34)

## V. MODEL CONVEXIFICATION

Note that **SRS-MINLP** contains several nonconvex terms, i.e., (10), (20), (21), (27), (28), (31), (33), and (34), making the model computationally intractable. In this section, the **SRS-MINLP** is transformed into a MISOCP model using several convex relaxation techniques. The exactness conditions are also represented.

*A. Relaxation of Nonconvex Constraints*

For the electricity distribution system, the branch flow definition (10) is a quadratic equality and hence nonconvex. It can be relaxed as a second-order cone constraint [27].

$$|S_{ij}|^2 \le v_i l_{ij}, i \to j \in \mathcal{E}^E \quad (35)$$

This relaxation is exact in radial networks and has been used in restoration problem [4].

For the water supply system, the hydraulic characteristics of pipes (20) is a quadratic equality and can be relaxed into a convex inequality [12].

$$\Delta h_{ij} \ge F_{ij}W_{ij}^2, i \to j \in \mathcal{E}_W^{pipe} \quad (36)$$

$$P_1 = \sum_{i\to j\in\mathcal{E}_W^{pipe}} \Delta h_{ij} \quad (37)$$

The inequality is tightened to an equality in radial networks with penalty term $P_1$ appended to the objective [12]. For meshed network, it is exact under some conditions: 1) all nodes in a ring have identical pressure limit; 2) all nodes in a ring but root node host no tanks or reservoirs; 3) all edges in a ring host no pump.

The hydraulic characteristic of water pumps (21) contains a logic proposition. To eliminate the *if* expression, the big-M method is used to rewrite the constraint.

$$-M(1-\chi_m) - (\alpha W_{ij} + \beta) \le \Delta h_{ij},$$
$$i \to j \in \mathcal{E}_W^{pump}, m \in \mathcal{N}^{pump} \quad (38)$$

$$\Delta h_{ij} \le M(1-\chi_m) - (\alpha W_{ij} + \beta),$$
$$i \to j \in \mathcal{E}_W^{pump}, m \in \mathcal{N}^{pump} \quad (39)$$

$$W_{ij} \le \chi_m M,$$
$$i \to j \in \mathcal{E}_W^{pump}, m \in \mathcal{N}^{pump} \quad (40)$$

where $M$ is a sufficiently large positive constant. When $\chi_m = 1$, the water pump is switched ON. Constraints (38) and (39) yield $\Delta h_{ij} = -(\alpha W_{ij} + \beta)$ and (40) holds trivially. When $\chi_m = 0$, the water pump is switched OFF. Constraints (38) and (39) hold trivially and (40) enforces $W_{ij} = 0$.

For the natural gas supply system, the pressure drop constraint (27) is a quadratic equality and the gas compressor operational constraint (28) contains a logical proposition. Similar to the constraints of the water supply system, they can be relaxed as

$$C_{ij}\Delta\psi_{ij} \ge G_{ij}^2, i \to j \in \mathcal{E}_G^{pipe} \quad (41)$$

$$-M(1-\chi_m) + \psi_i \le \psi_j, \quad (42)$$

$$i \rightarrow j \in \mathcal{E}_G^{\text{comp}}, m \in \mathcal{N}^{\text{comp}}$$

$$\psi_j \leq M(1 - \chi_m) + \gamma \psi_i,$$
$$i \rightarrow j \in \mathcal{E}_G^{\text{comp}}, m \in \mathcal{N}^{\text{comp}} \quad (43)$$

$$G_{ij} \leq \chi_m M, \quad i \rightarrow j \in \mathcal{E}_G^{\text{comp}}, m \in \mathcal{N}^{\text{comp}} \quad (44)$$

Constraint (41) is tightened to an equality in the radial network with penalty terms $P_2$ appended to the objective [13].

$$P_2 = \sum_{i \rightarrow j \in \mathcal{E}_G^{\text{pipe}}} \Delta \psi_{ij} \quad (45)$$

For water pumps, the electricity consumption is a quadratic equality (31) and can be relaxed into an inequality.

$$P_m^{\text{pump}} \geq \frac{\rho^W g^W}{\eta^{\text{pump}}} (\alpha W_{ij}^2 + \beta W_{ij}),$$
$$m \in \mathcal{N}^{\text{pump}}, i \rightarrow j \in \mathcal{E}_W^{\text{pump}} \quad (46)$$

The relaxation is exact if the following conditions are satisfied.

**Condition 1**: Not all water demands are restored.

**Condition 2**: The water flow on pipes does not reach its upper bound.

**Condition 3**: The water head at reservoir node is greater than the sum of water head losses on pipes under maximum allowable water flow, i.e., $h_1 \geq \sum_{i \rightarrow j \in \mathcal{E}_W^{\text{pipe}}} F_{ij} \overline{W_{ij}}^2$.

The proof is given in Appendix A.

For nodes with water pumps connected, the injected complex power (33) is determined by ON/OFF status and electricity consumption. The bilinear term $\chi_i P_i^{\text{pump}}$ makes constraint (33) nonconvex. This constraint can be equivalently expressed via four linear inequalities using McCormick linearization method [33].

$$-\chi_i \left( \overline{P_i^{\text{pump}}} + j\varphi_i^{\text{pump}} \overline{P_i^{\text{pump}}} \right) \leq s_i, i \in \mathcal{N}^{\text{pump}} \quad (47)$$

$$-\left( P_i^{\text{pump}} + j\varphi_i^{\text{pump}} P_i^{\text{pump}} \right) \leq s_i, i \in \mathcal{N}^{\text{pump}} \quad (48)$$

$$s_i \leq 0, i \in \mathcal{N}^{\text{pump}} \quad (49)$$

$$s_i \leq -\left( P_i^{\text{pump}} + j\varphi_i^{\text{pump}} P_i^{\text{pump}} \right) + (1 - \chi_i) \left( \overline{P_i^{\text{pump}}} + j\varphi_i^{\text{pump}} \overline{P_i^{\text{pump}}} \right), i \in \mathcal{N}^{\text{pump}} \quad (50)$$

Note that when $\chi_i = 1$, constraints (47)-(50) yields $s_i = -(P_i^{\text{pump}} + j\varphi_i^{\text{pump}} P_i^{\text{pump}})$. When $\chi_i = 0$, constraints (47)-(50) enforce $s_i = 0$.

For nodes with gas compressors connected, the injected complex power (34) can be replaced by constraints (51)-(54).

$$-\chi_i \left( \overline{P_i^{\text{comp}}} + j\varphi_i^{\text{comp}} \overline{P_i^{\text{comp}}} \right) \leq s_i, i \in \mathcal{N}^{\text{comp}} \quad (51)$$

$$-\left( P_i^{\text{comp}} + j\varphi_i^{\text{comp}} P_i^{\text{comp}} \right) \leq s_i, i \in \mathcal{N}^{\text{comp}} \quad (52)$$

$$s_i \leq 0, i \in \mathcal{N}^{\text{comp}} \quad (53)$$

$$s_i \leq -\left( P_i^{\text{comp}} + j\varphi_i^{\text{comp}} P_i^{\text{comp}} \right) + (1 - \chi_i) \left( \overline{P_i^{\text{comp}}} + j\varphi_i^{\text{comp}} \overline{P_i^{\text{comp}}} \right), i \in \mathcal{N}^{\text{comp}} \quad (54)$$

### B. Convex Model

As a result, a MISOCP model is formulated.

**SRS-MISOCP**

$$\max \quad f - \lambda(P_1 + P_2)$$
$$\text{over:} \quad a_{ij}, F_{ij}, r_i, w_i, g_i, P_i^{\text{pump}}, P_i^{\text{comp}}, s_i^{\text{DG}}, S_{ij}, s_i,$$
$$v_i, l_{ij}, \chi_i, w_i^G, W_{ij}, h_i, \Delta h_{ij}, g_i^G, G_{ij}, \psi_i, \Delta \psi_{ij}$$
$$\text{s.t.} \quad (5)\text{-}(9), \ (11)\text{-}(19), \ (22)\text{-}(26), \ (29)\text{-}(30), \ (32),$$
$$(35)\text{-}(36), \ (38)\text{-}(44), \ (46)\text{-}(54)$$

## VI. CASE STUDIES

The proposed method is applied to two cases. The first one is to show the service restoration strategy and the benefit of considering interdependency. The second one is to validate the exactness and effectiveness of the proposed method.

### A. Tests on the Benefit of Considering Interdependency

In this part, an electricity-water-gas integrated system is developed and simulated to show the benefits of modeling interdependency.

#### 1) Case 1: Test System Information

In this part, an integrated electricity-water-gas system is proposed that consists of a 32-node electricity system [26], a 15-node water system [31], and a 20-node natural gas supply system [20]. The topologies of three subsystems are shown in Fig. 3. Customer IDs are marked inside the pentagram, square, and pentagon symbols in the diagrams of the three subnetworks, indicating their connections on the customer side. Note that pumps in a water supply system and compressors in a natural gas supply system may be served by multiple feeders in the distribution system. The test system only includes one feeder for simplification. The proposed method can be readily extended to systems with multiple feeders.

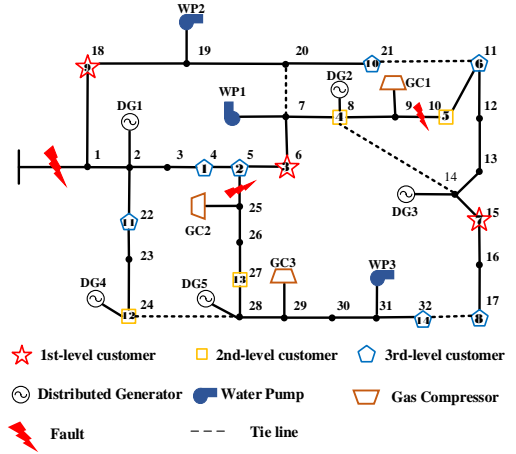

(a) Diagram of the 32-node electricity distribution network

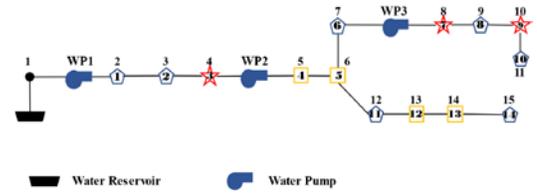

(b) Diagram of the 15-node water supply network

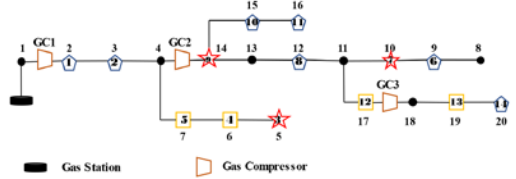

(c) Diagram of the 20-node natural gas supply network

Fig. 3. Diagrams of electricity-water-gas integrated system for Case 1.

Three water pumps and three gas compressors are connected to the electricity subsystem. The detailed information is given in Table I, [34], [20]. Especially, abbreviations WP and GC in Table I represent water pump and gas compressor, respectively.

TABLE I
WATER PUMPS AND GAS COMPRESSORS INFORMATION

| Type | Water/Gas Branch | Electricity Node | Rated Active Power (kW) | Parameters |
|------|------------------|------------------|-------------------------|------------|
| WP1 | 1-2 | 7 | 350 | $\alpha = 185$ |
| WP2 | 4-5 | 19 | 250 | $\beta = 223.32$ |
| WP3 | 7-8 | 31 | 150 | $\eta^{pump} = 0.8075$ |
|      |     |    |     | $\varphi^{pump} = 0.85$ |
| GC1 | 1-2 | 9 | 350 | $\gamma = 4$ |
| GC2 | 4-14 | 25 | 250 | $\sigma = 0.42$ |
| GC3 | 17-18 | 29 | 150 | $\varphi^{comp} = 0.85$ |

Fourteen customers in the integrated system are divided into three levels with the priority weights 100, 10, and 0.2, respectively [3]. Detailed information of the subsystems is given in Appendix B. The weighting factors $a_i$, $b_i$, and $c_i$ of all customers are set as 0.4, 0.3, and 0.3. Five DGs provide electricity, whose capacities are 740, 330, 440, 460, and 230 kVA. Weighting coefficients $\lambda$ are set as 0.001.

*2) Service Restoration Strategy*

Suppose that after an extreme event, the faults happened on 9-10 and 5-25, the electricity distribution system is disconnected from the main grid. The supply of electricity, water, and natural gas to fourteen customers is interrupted. The water pumps and gas compressors are switched OFF. The model is solved using the optimization toolbox YALMIP [35] along with the solver GUROBI, a commercial solver suitable for general NLP and MINLP [36]. The restoration status of customers is shown in Fig. 4.

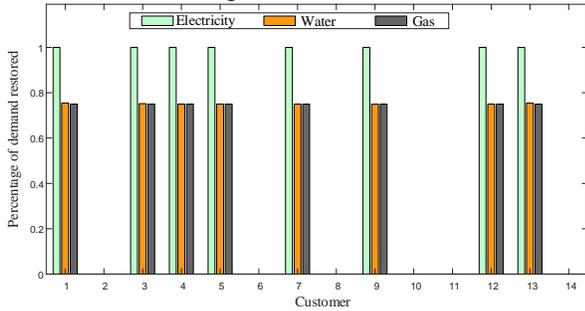

Fig. 4. Restoration status of electricity, water, and gas demand of each customer.

As shown in Fig. 4, the ratio of restored electricity, water, and natural gas demand are close to the ratio of $a_i$, $b_i$, and $c_i$. The switches 7-20, 8-14, 11-21, and 17-32 are closed, and the switches 2-3 and 28-29 are opened. All 1st-level, 2nd-level, and one 3rd-level customers are restored. All water pumps and gas compressors are switched ON. The voltage magnitude profile in the electricity distribution system is shown in Fig. 5.

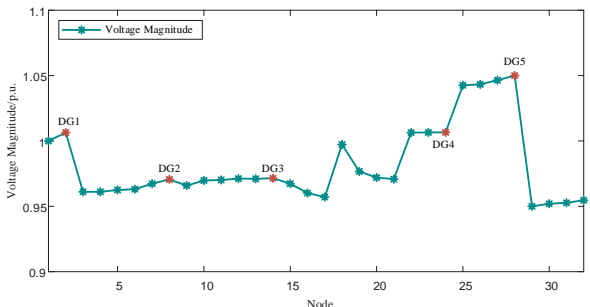

Fig. 5. Voltage profile in the electricity distribution system after restoration.

According to Fig. 5, the voltage magnitude at each node is kept in the feasible range. The total electricity consumption of water pumps and gas compressors is 637.71 kW, while the restored electricity demand of customers is 995 kW. Nearly 40% of electricity is allocated to water pumps and gas compressors to restore water and natural gas supplies to customers.

*3) The Benefit of Considering Interdependency*

In this part, the benefit of considering interdependency is analyzed by comparing with the restoration results from the existing restoration method (i.e., **CLR-misocp** proposed in [26]), in which the restoration status of all electricity loads including water pumps and gas compressors is first determined, then the restoration states of water and natural gas supply systems are determined independently. The objective is to maximize the weighted sum of restored loads, including customers (number, described as (2)), water pumps (percentage, described as (3)), and gas compressors (percentage, described as (3)). The priority weights settings are listed in Table II. The restoration results of the proposed method and existing restoration method **CLR-misocp** are shown in Table III.

TABLE II
PRIORITY WEIGHTS SETTINGS OF PROPOSED METHOD AND CLR-MISOCP METHOD

|  | Proposed method | CLR-misocp method | |
|--|-----------------|-----------|-----------|
|  |  | Setting 1 | Setting 2 |
| Water Pumps | - | 10000 | 100 |
| Gas Compressors | - | 10000 | 100 |
| 1st-level customer | 100 | 100 | 100 |
| 2nd-level customer | 10 | 10 | 10 |
| 3rd-level customer | 0.2 | 0.2 | 0.2 |

TABLE III
RESTORATION RESULTS OF PROPOSED METHOD AND CLR-MISOCP METHOD

|  | Proposed method | CLR-misocp method | |
|--|-----------------|-----------|-----------|
|  |  | Setting 1 | Setting 2 |
| Number of restored electricity customers | 11 | 2 | 3 |
| Percentage of restored water demand (%) | 47.23% | 100% | 97.57% |
| Percentage of restored gas demand (%) | 45.65% | 100% | 69.29% |
| Restored WPs and GCs | WP1 (44.91%), WP2 (43.17%), WP3 (49.59%) GC1 (41.53%), GC2 (38.44%), GC3 (37.84%) | WP1 (100%), WP2 (100%), WP3 (100%) GC1 (100%), GC2 (100%), GC3 (100%) | WP1 (82.75%), WP2 (100%), WP3 (100%) GC1 (63.04%), GC2 (100%), GC3 (100%) |

For Setting 1, water and natural gas demand of all 14 customers are fully satisfied while only the electricity demand of customers 7 and 9 are restored. The electricity, water, and natural gas supplies to most customers cannot maintain their critical function.

For Setting 2, the result is optimal for objective value improvement but not optimal for restoring the function of water and natural gas systems, neither for the customers. In **CLR-misocp** method, the loads that have higher priorities are more likely to be restored. For the loads that have the same priority weights, the ones with less amount of active power demand will be restored first since the restoration of per kW of them increases the objective value more. Therefore, the upstream water pumps and gas compressors (WP1 and GC1) are partially restored while the downstream ones with less amount are fully restored, resulting in waste of generation resources. Taking the path 1-2-3-4-14-15-16 of the natural gas subsystem as an example. The natural gas demand of customers 15 and 16 are not restored due to the limited gas transferred by the partial function of GC1. In contrast, the actual electricity consumption of GC2 is less than the allocated power. A better strategy may be to allocate more electricity to GC1 while assigning less to GC2.

### 4) Discussions on Weighting Factors $a_i$, $b_i$, $c_i$

In this part, the effect of weighting factors $a_i$, $b_i$, and $c_i$ in the objective function is discussed. The restoration results of four cases with different $a_i$, $b_i$, and $c_i$ are shown in Table IV. $\{a_i, b_i, c_i\}$ is set as $\{0.3, 0.3, 0.3\}$, $\{0.8, 0.1, 0.1\}$, $\{0.1, 0.8, 0.1\}$, and $\{0.1, 0.1, 0.8\}$, respectively.

TABLE IV
ALLOCATION OF ELECTRICITY WITH DIFFERENT SETTINGS OF $\{a_i, b_i, c_i\}$

| | Case A | Case B | Case C | Case D |
|---|---|---|---|---|
| Electricity allocation to customers (%) | 56.17 | **90.29** | 66.91 | 70.14 |
| Electricity allocation to WPs (%) | 24.84 | 4.97 | **29.75** | 3.75 |
| Electricity allocation to GCs (%) | 18.98 | 4.74 | 3.34 | **26.12** |

According to Table IV, more generation resource is allocated to electricity demands when $a_i$ increases. The ratio of electricity allocated to water pumps and gas compressors is close to the ratio of $b$ and $c$. It shows that the trend of generation resource allocation is affected by $a_i$, $b_i$, and $c_i$.

### B. Tests on the Exactness and Computation Efficiency

In this part, the proposed method is applied to a large-scale case to validate the exactness and computational efficiency of the convex relaxation.

#### 1) Case 2: Test System Information

An integrated system is developed by combining a 123-node electric power distribution feeder [3], a 42-node water supply system [37], and a 40-node natural gas supply system. The diagrams of the electricity, water, and natural gas subsystems are shown in Fig. 6 (a), (b), and (c), respectively. The natural gas supply system shown in the Fig. 3 (c) is used in this case and two customers are connected at each node with a short branch. The detailed information is shown in Appendix C.

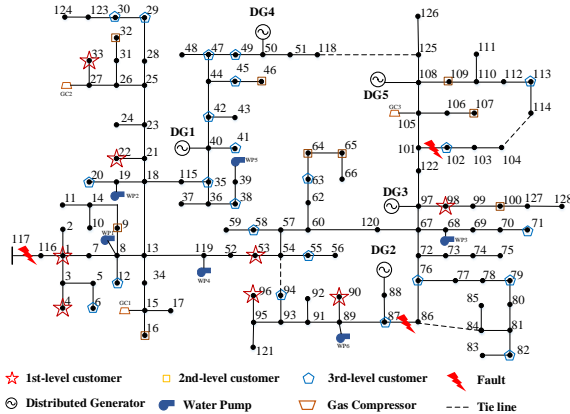

(a) Diagram of the 123-node electricity distribution network

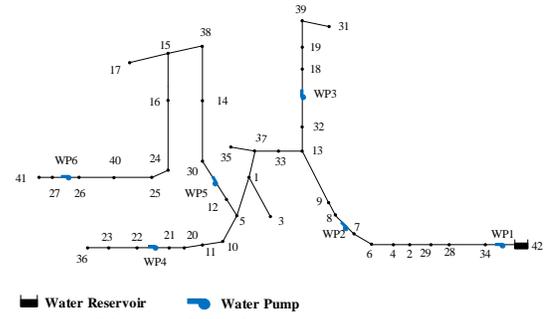

(b) Diagram of the 42-node water supply network

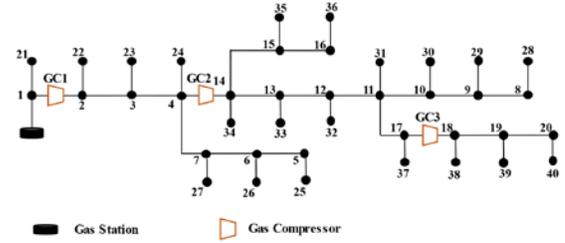

(c) Diagram of the 40-node natural gas supply network

Fig. 6. Diagrams of the large-scale electricity-water-gas integrated system for Case 2.

#### 2) Exactness of Convex Relaxation

In this part, the exactness of proposed method is evaluated by checking the exactness gap of the relaxed constraint (46) and comparing the restoration results of **SRS-MINLP** and **SRS-MISOCP**. The exactness gap $R$ is defined as

$$R = \max_{\substack{m \in \mathcal{N}^{\text{pump}}, \\ i \to j \in \mathcal{E}_W^{\text{pump}}}} \frac{P_m^{\text{pump}} - \frac{\rho^W g^W}{\eta^{\text{pump}}} \left( \alpha W_{ij}^2 + \beta W_{ij} \right)}{\frac{\rho^W g^W}{\eta^{\text{pump}}} \left( \alpha W_{ij}^2 + \beta W_{ij} \right)} \quad (55)$$

The two models are both solved by GUROBI. We generated 20 scenarios with different DG capacities, customer demands, and critical load locations. The value of $R$ is less than $10^{-4}$ in 20 scenarios. For results of two models, the values of integer variables of 20 scenarios are the same. The deviation of two optimums does not exceed $10^{-3}$, indicating that the relaxation is exact.

#### 3) Computation Efficiency

In this part, the computation efficiency of proposed convex relaxation method is tested. The computation time of **SRS-MINLP** and **SRS-MISOCP** are listed in Table V.

TABLE V
THE COMPUTATION TIME OF DIFFERENT METHODS

| Program | Computation Time (s) | | | Number of scenarios exceeding 30 min |
|---|---|---|---|---|
| | Ave. | Min. | Max. | |
| SRS-MINLP | 2473.63 | 89.51 | 15217.91 | 7 |
| SRS-MISOCP | 104.99 | 13.69 | 371.96 | 0 |

According to Table V, in 7 scenarios, the computation time of solving **SRS-MINLP** exceeds 30 minutes, indicating it cannot meet the online decision requirement [29]. The **SRS-MISOCP** speeds up the solution process by 23.56 times in average time compared with **SRS-MINLP**. To further demonstrate the efficiency, we solve the **SRS-MISOCP** in 1000 scenarios. The distribution of computation time for 1000 scenarios is shown in Fig. 7.

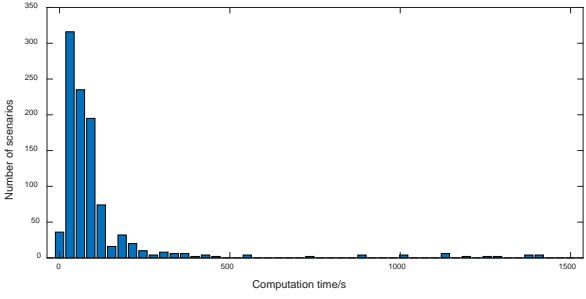

Fig. 7. The distribution of computation time for 1000 scenarios.

As shown in Fig. 7, the computation time of **SRS-MISOCP** in all scenarios is less than 30 minutes. The average computation time is 115.51s and the maximum is 1416.53s. The solving process can be completed in 10 minutes for more than 950 scenarios. For scenarios with longer computation time, the DG capacity is relatively large which can ensure all 1st- and 2nd-level customers are restored, indicating the restoration status of 3rd-level customers are the true decision variables. However, many 3rd-level customers have the same demand thus have the similar little effect on the objective, which makes the mixed-integer solution algorithm (such as branch-and-cut algorithm) converge slowly to find a better solution. Even so, the computation time of **SRS-MISOCP** under these scenarios is still less than **SRS-MINLP** method. The result shows that the proposed method has a high computational performance and is promising as an online decision support tool.

## VII. CONCLUSION

In this paper, a service restoration method is proposed for electricity distribution systems considering the electricity-water-gas interdependency. A MISOCP model is developed using relaxation techniques and the exactness conditions are discussed. The case studies indicate that the proposed model can be used to allocate limited generation resources among electricity, water, and natural gas supply systems to keep the critical function of customers after restoration.

This paper focuses on the dependence of water pumps and gas compressors on electricity in the distribution level. In the future work, more coupling components, such as water treatment plants, water cooling towers [38], and gas-fired DGs, should be incorporated in the service restoration model. Interdependency of electricity, water, and natural gas supply systems in larger scales, e.g., the transmission level [39], should also be studied.

## APPENDIX A
### PROOF OF RELAXATION EXACTNESS

To prove that the relaxation is exact, it suffices to show that any optimal solution of **SRS-MISOCP** attains equality in (46). Assume that $\mathcal{S}_{cr} = \{r_i, w_i, g_i, s_i^{DG}, P_m^{pump}, P_m^{comp}, W_{ij}\}$ is the set of optimal solution for **SRS-MISOCP**, but a water supply system branch $i \to j \in \mathcal{E}_W^{pump}$ has strict inequality, i.e., $P_m^{pump} > \frac{\rho^W g^W}{\eta^{pump}}(\alpha W_{ij}^2 + \beta W_{ij})$.

According to the objective, only the minimum of $\frac{r_i}{a_i}, \frac{w_i/w_i^{max}}{b_i}$, and $\frac{g_i/g_i^{max}}{c_i}$ contributes the objective value. Since $r_i$ is discrete, as long as not all $r_i = 0$, the objective will be decided by the value of $\frac{w_i/w_i^{max}}{b_i}$ or $\frac{g_i/g_i^{max}}{c_i}$ indeed.

If $\frac{g_i/g_i^{max}}{c_i} \leq \frac{w_i/w_i^{max}}{b_i}$, the objective value is determined by $\frac{g_i/g_i^{max}}{c_i}$. If (46) is a strictly inequality, it means some electricity is allocated to water pumps but do not contribute to their function. These portions of electricity can be reallocated to gas compressors to transfer more resource and further increase the objective value, which contradicts the assumption that it is the optimal solution.

If $\frac{g_i/g_i^{max}}{c_i} > \frac{w_i/w_i^{max}}{b_i}$, the objective value is determined by $\frac{w_i/w_i^{max}}{b_i}$. Introduce $\xi > 0$ and consider another set $\mathcal{S}_{cr}^* = \{r_i^*, w_i^*, g_i^*, s_i^{DG*}, P_m^{pump*}, P_m^{comp*}, W_{ij}^*\}$ denoted by

$$r_i^* = r_i \qquad w_j^* = w_j + \xi \qquad g_i^* = g_i$$
$$s_i^{DG*} = s_i^{DG} \qquad P_m^{pump*} = P_m^{pump} \qquad P_m^{comp*} = P_m^{comp}$$
$$W_{\Theta(j)}^* = W_{\Theta(j)} \qquad W_{\Omega(j)}^* = W_{\Omega(j)} + \xi \qquad h_1^* = h_1$$

where $\Omega(j)$ means the upstream branches of node $j$ and $\Theta(j)$ means the downstream branches. Since $w_j^* > w_j$, $\mathcal{S}_{cr}^*$ has a strictly larger objective value than $\mathcal{S}_{cr}$. If $\mathcal{S}_{cr}^*$ is a feasible set, then it nullifies the optimality of $\mathcal{S}_{cr}$.

It suffices then to check that there exists $\xi > 0$ that $\mathcal{S}_{cr}^*$ satisfies **SRS-MISOCP** constraints. Since $\mathcal{S}_{cr}$ is feasible, despite water supply system operational constraints (17)-(19), (22)-(23), (36), (38)-(40), and (46), the others hold for $\mathcal{S}_{cr}^*$ as well. Under condition 1 and condition 2, constraints (17)-(19), (22), (36), and (38)-(40) hold for $\mathcal{S}_{cr}^*$. To prove that constraint (23) is satisfied, for each node $k$ incident to $\Omega(j)$, assign the water head consistent with (20) along the path from water reservoir to $k$

$$h_k^* = h_1^* + \sum_{i \to j \in \Omega(k)} \alpha W_{ij}^* + \beta - \sum_{i \to j \in \Omega(k)} F_{ij} W_{ij}^{*2}$$

Under condition 3, $h_k^*$ is larger than the allowable water head. For nodes incident to $\Theta(j)$, $h_k^* = h_k$ and constraint (23) is satisfied. For constraint (46) across $i \to j$, it is obtained that

$$P_m^{pump*} - \frac{\rho^W g^W}{\eta^{pump}}(\alpha W_{ij}^{*2} + \beta W_{ij}^*)$$
$$= P_m^{pump} - \frac{\rho^W g^W}{\eta^{pump}}\left(\alpha(W_{ij} + \xi)^2 + \beta(W_{ij} + \xi)\right)$$
$$= P_m^{pump} - \frac{\rho^W g^W}{\eta^{pump}}(\alpha W_{ij}^2 + \beta W_{ij}) - \frac{\rho^W g^W}{\eta^{pump}}(2\alpha W_{ij}\xi + \alpha\xi^2 + \beta\xi)$$

Since $P_m^{pump} - \frac{\rho^W g^W}{\eta^{pump}}(\alpha W_{ij}^2 + \beta W_{ij}) > 0$, we can choose a $\xi > 0$ sufficiently small that $P_m^{pump} - \frac{\rho^W g^W}{\eta^{pump}}(\alpha W_{ij}^{*2} + \beta W_{ij}^*) > 0$. This completes the proof.

## APPENDIX B
### DETAILED INFORMATION OF CASE 1

The node and branch information of the electricity subsystem is given in [26]. The water subsystem is a modified 15-node water distribution test system. The node data are listed in Table B1 and the branch data in [31]. For natural gas subsystem, the node data are illustrated in Table B2 and the branch data are available in [20].

TABLE B1
THE NODE DATA OF WATER SUBSYSTEM IN CASE 1

| Node | Demand ($m^3/h$) | Minimum pressure (m) | Node | Demand ($m^3/h$) | Minimum pressure (m) |
|---|---|---|---|---|---|
| 1 | 0 | 10.6 | 9 | 11.88 | 177 |
| 2 | 30.06 | 29.7 | 10 | 119.16 | 176 |
| 3 | 40.32 | 20 | 11 | 10.8 | 176.65 |
| 4 | 49.32 | 25 | 12 | 14.4 | 148 |
| 5 | 7.2 | 75 | 13 | 14.4 | 144.13 |
| 6 | 3.6 | 146 | 14 | 30.06 | 143 |
| 7 | 5.04 | 141 | 15 | 72 | 126.42 |
| 8 | 7.2 | 146 | | | |

TABLE B2
THE NODE DATA OF NATURAL GAS SUBSYSTEM IN CASE 1

| Node | Demand ($m^3/h$) | Node | Demand ($m^3/h$) | Node | Demand ($m^3/h$) |
|---|---|---|---|---|---|
| 1 | 0 | 8 | 0 | 15 | 88.8 |
| 2 | 30 | 9 | 36.7 | 16 | 13.2 |
| 3 | 51 | 10 | 54.1 | 17 | 33.4 |
| 4 | 0 | 11 | 0 | 18 | 0 |
| 5 | 52.2 | 12 | 82.4 | 19 | 180.2 |
| 6 | 41.1 | 13 | 0 | 20 | 24.6 |
| 7 | 33.1 | 14 | 37.4 | | |

APPENDIX C
*DETAILED INFORMATION OF CASE 2*

In Case 2, six water pumps and three gas compressors with same rated active power, 500kW, are connected. The other parameters of them can be found in Table I.

Forty customers in the integrated system are divided into three levels. The location of customers at electricity and water subsystems is listed in Table C1. The customer IDs are same as the node IDs in the natural gas subsystem. The node and branch information of the electricity subsystem is given in [3]. The node data of the water subsystem are listed in Table C2 and the branch data are given in [37]. For natural gas subsystem, the node data are illustrated in Table C3.

The capacities of five DGs are 800, 900, 950, 730, and 630 kVA, respectively. The priority weights of customers and the weighting factors of objective are same as those in Case 1.

TABLE C1
THE LOCATION OF FORTY CUSTOMERS AT ELECTRICITY AND WATER SUBSYSTEMS IN CASE 2

| Customer | Electricity | Water | Customer | Electricity | Water |
|---|---|---|---|---|---|
| 1 | 1 | 1 | 21 | 53 | 22 |
| 2 | 4 | 2 | 22 | 55 | 23 |
| 3 | 6 | 3 | 23 | 58 | 24 |
| 4 | 9 | 4 | 24 | 63 | 25 |
| 5 | 12 | 6 | 25 | 64 | 26 |
| 6 | 16 | 7 | 26 | 65 | 27 |
| 7 | 20 | 8 | 27 | 71 | 28 |
| 8 | 22 | 9 | 28 | 76 | 29 |
| 9 | 29 | 10 | 29 | 79 | 30 |
| 10 | 30 | 11 | 30 | 82 | 31 |
| 11 | 32 | 12 | 31 | 87 | 32 |
| 12 | 33 | 13 | 32 | 90 | 33 |
| 13 | 35 | 14 | 33 | 94 | 34 |
| 14 | 38 | 15 | 34 | 96 | 35 |
| 15 | 41 | 16 | 35 | 98 | 36 |
| 16 | 42 | 17 | 36 | 100 | 37 |
| 17 | 45 | 18 | 37 | 102 | 38 |
| 18 | 46 | 19 | 38 | 107 | 39 |
| 19 | 47 | 20 | 39 | 109 | 40 |
| 20 | 49 | 21 | 40 | 113 | 41 |

TABLE C2
THE NODE DATA OF WATER SUBSYSTEM IN CASE 2

| Node | Demand ($m^3/h$) | Minimum pressure (m) | Node | Demand ($m^3/h$) | Minimum pressure (m) |
|---|---|---|---|---|---|
| 1 | 7.2 | 183 | 22 | 5.4 | 139.82 |
| 2 | 3.6 | 68.85 | 23 | 3.6 | 140 |
| 3 | 3.6 | 166.42 | 24 | 3.96 | 198.33 |
| 4 | 14.4 | 60 | 25 | 3.6 | 202.15 |
| 5 | 0 | 184 | 26 | 4.68 | 177 |
| 6 | 11.916 | 65 | 27 | 3.6 | 177 |
| 7 | 4.32 | 65 | 28 | 18 | 69.39 |
| 8 | 4.32 | 112.18 | 29 | 11.52 | 69 |
| 9 | 6.12 | 115 | 30 | 14.4 | 184 |
| 10 | 5.4 | 134.99 | 31 | 14.4 | 216.65 |
| 11 | 14.4 | 101 | 32 | 14.4 | 186 |
| 12 | 14.4 | 184 | 33 | 4.32 | 188 |
| 13 | 3.6 | 186 | 34 | 5.4 | 69 |
| 14 | 14.4 | 242 | 35 | 3.6 | 100 |
| 15 | 3.6 | 242 | 36 | 46.8 | 258.9 |
| 16 | 3.6 | 242 | 37 | 4.68 | 184.13 |
| 17 | 3.96 | 242 | 38 | 4.68 | 241.18 |
| 18 | 3.96 | 185.78 | 39 | 6.84 | 240 |
| 19 | 6.12 | 217 | 40 | 3.96 | 203.01 |
| 20 | 6.12 | 139.02 | 41 | 39.6 | 235.71 |
| 21 | 5.76 | 139.82 | 42 | 0 | 1 |

TABLE C3
THE NODE DATA OF NATURAL GAS SUBSYSTEM IN CASE 2

| Node | Demand ($m^3/h$) | Node | Demand ($m^3/h$) | Node | Demand ($m^3/h$) |
|---|---|---|---|---|---|
| 1 | 15 | 15 | 3.31 | 29 | 36.7 |
| 2 | 10 | 16 | 6.6 | 30 | 5.51 |
| 3 | 21 | 17 | 33.4 | 31 | 35.2 |
| 4 | 22.2 | 18 | 33.2 | 32 | 15.6 |
| 5 | 12.1 | 19 | 20.1 | 33 | 18.4 |
| 6 | 33.21 | 20 | 20 | 34 | 19.6 |
| 7 | 11.2 | 21 | 31.2 | 35 | 17.3 |
| 8 | 2.5 | 22 | 13.4 | 36 | 36.2 |
| 9 | 16 | 23 | 21 | 37 | 22.1 |
| 10 | 27.1 | 24 | 15.3 | 38 | 24.6 |
| 11 | 31.2 | 25 | 28.2 | 39 | 48.2 |
| 12 | 42.4 | 26 | 3.74 | 40 | 24.6 |
| 13 | 19.3 | 27 | 11.9 | | |
| 14 | 10.2 | 28 | 18.9 | | |

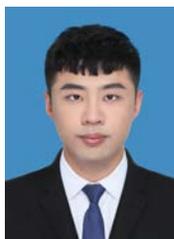

**Jiaxu Li** (S'18) received the B.E. degree from Nanchang University, Nanchang, China, in 2017 and the M.S. degree in Electrical Engineering from Beijing Jiaotong University, Beijing, China, in 2019. He is pursuing the Ph.D. degree in Electrical Engineering at Beijing Jiaotong University.

His research interests include power system resilience and distribution system restoration considering critical infrastructure interdependency.

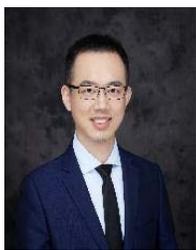

**Yin Xu** (S'12–M'14–SM'18) received the B.E. and Ph.D. degrees in electrical engineering from Tsinghua University, Beijing, China, in 2008 and 2013, respectively. From 2013 to 2016, he was an Assistant Research Professor with the School of Electrical Engineering and Computer Science, Washington State University, Pullman, WA, USA. He is currently a Professor with the School of Electrical Engineering, Beijing Jiaotong University, Beijing, China. His research interests include power grid resilience, transportation electrification, and power system electromagnetic transient modeling and high-performance simulation.

Dr. Xu is currently serving as Chair of the Energy Internet Resilience Working Group under the IEEE PES Energy Internet Coordinating Committee and Secretary of the Distribution Test Feeder Working Group under the IEEE PES Distribution System Analysis Subcommittee. He is an Editorial Broad member of IEEE Transactions on Power Systems, IEEE Power Engineering Letters, IET Smart Grid, and IET Energy Conversion and Economics.

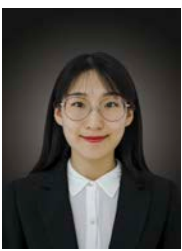

**Ying Wang** (S'16–M'20) received the B.E. and Ph.D. degrees in Electrical Engineering from Beijing Jiaotong University, Beijing, China, in 2014 and 2020, respectively. From 2017 to 2018, she was a visiting scholar at Case Western Reserve University, Cleveland, OH, USA. She is currently a Post-Doctoral Researcher at Beijing Jiaotong University, Beijing, China.

Her research interests include power system resilience, distribution system restoration, and restoration for the distribution system coupled with the transportation system.

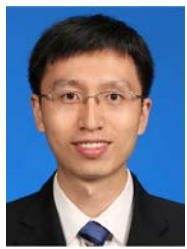

**Meng Li** (M'18) received the B.S. and Ph.D. degrees in electrical engineering from North China Electric Power University, Beijing, China, in 2003 and 2018, respectively. He is currently an associate professor of Electrical Engineering with the School of Electrical Engineering, Beijing Jiaotong University, Beijing, China.

His research interests include dc grid protection and power system resilience.

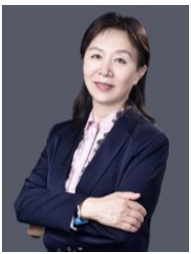

**Jinghan He** (M'07–SM'18–F'20) received her M.Sc. degree in automation from Tianjin University, Tianjin, China, in 1994 and the Ph.D. degree in electrical engineering from Beijing Jiaotong University, Beijing, China, in 2007. She has been a professor with the School of Electrical Engineering, Beijing Jiaotong University since 2000.

Her research interests include power system relay protection, monitoring and protection of the railway traction power supply systems, DC grid and hybrid DC grid protection and control, renewable energy and smart grid.

Prof. He is currently serving as a Chairman of IEEE PES China Membership Committee and a member of DC System Protection and Control Working Group under the IEEE PES China Subcommittee.

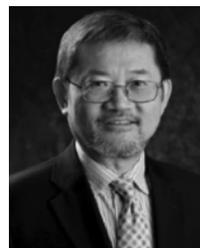

**Chen-Ching Liu** (S'80–M'83–SM'90–F'94–LF) received the Ph.D. degree from the University of California, Berkeley, CA, USA, in 1983, and the Doctor Honoris Causa degree from the Polytechnic University of Bucharest, Romania, in 2013.

He served as a Professor with the University of Washington, Seattle, WA, USA, from 1983 to 2005. From 2006 to 2008, he was a Palmer Chair Professor with Iowa State University, Ames, IA, USA. From 2008 to 2011, he was a Professor and an Acting/Deputy Principal of the College of Engineering, Mathematical and Physical Sciences, University College Dublin, Ireland. He was a Boeing Distinguished Professor of electrical engineering and the Director of the Energy Systems Innovation Center of Washington State University, Pullman, WA, USA. He is currently an American Electric Power Professor and the Director of the Power and Energy Center of Virginia Polytechnic Institute and State University, Blacksburg, VA, USA. He is also a Research Professor with Washington State University.

Prof. Liu was a recipient of the IEEE PES Outstanding Power Engineering Educator Award in 2004. He served as the Chair of the IEEE PES Technical Committee on Power System Analysis, Computing, and Economics from 2005 to 2006. He is the U.S. Member of CIGRE Study Committee D2, Information Systems and Telecommunication. He is also a Member of the U.S. National Academy of Engineering.

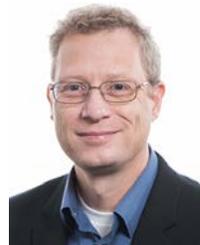

**Kevin P. Schneider** (S'00–M'06–SM'08–F'20) received his B.S. degree in Physics and his M.S. and Ph.D. degrees in Electrical Engineering from the University of Washington. His main areas of research are distribution system analysis and power system operations.

He is currently a Laboratory Fellow at the Pacific Northwest National Laboratory, Manager of the Distribution and Demand Response Sub-Sector, and a Research Professor at Washington State University as part of the PNNL/WSU Advanced Grid Institute (AGI).

Dr. Schneider is an Affiliate Associate Professor at the University of Washington and a licensed Professional Engineer in Washington State. He is a Fellow of the IEEE, past chair of the Power & Energy Society (PES) Distribution System Analysis (DSA) Sub-Committee, and the past Chair of the Analytic Methods for Power Systems (AMPS) Committee.